\documentclass[epjST]{svjour}  
\usepackage{xcolor}  
\usepackage[colorlinks]{hyperref}   
\usepackage{appendix}
\usepackage{rotating, graphicx}
\usepackage{graphics}      
\usepackage{latexsym}
\begin{document}
\title{A Class of Isochronous and Non-Isochronous Nonlinear Oscillators}
\author{J. Ramya Parkavi\inst{1} \thanks{\emph{E-mail:} ramyaparkji@gmail.com } \and R. Mohanasubha\inst{2} \thanks{\emph{E-mail:}mohanasubhar@citchennai.net} \and V. K. Chandrasekar\inst{3} \thanks{\emph{E-mail:}chandru25nld@gmail.com (Corresponding Author)} \and M. Senthilvelan\inst{4} \thanks{\emph{E-mail:}senv0000@gmail.com} \and M. Lakshmanan\inst{5}
\thanks{\emph{E-mail:}lakshman.cnld@gmail.com}
 }                 
%
%
\institute{Centre for Nonlinear Science and Engineering, School of Electrical and Electronics Engineering, SASTRA Deemed University, Thanjavur-613401, Tamil Nadu, India \and Centre for Nonlinear Systems, Chennai Institute of Technology, Chennai-600069, Tamil Nadu, India \and Centre for Nonlinear Science and Engineering, School of Electrical and Electronics Engineering, SASTRA University, Thanjavur-613401, Tamil Nadu, India \and Department of Nonlinear Dynamics, School of Physics, Bharathidasan University, Tiruchirappalli - 620024, Tamil Nadu, India \and Department of Nonlinear Dynamics, School of Physics, Bharathidasan University, Tiruchirappalli - 620024, Tamil Nadu, India}
\date{Received: date / Revised version: date}
%
\abstract{
In this work, we present a method of generating a class of nonlinear ordinary differential equations (ODEs), representing the dynamics of appropriate nonlinear oscillators, that have the characteristics of either amplitude independent frequency of oscillations or amplitude dependent frequency of oscillations from the integrals of the simple harmonic oscillator equation. To achieve this, we consider the case where the integrals are in the same form both for the linear and the nonlinear oscillators in either of the cases. We also discuss the method of deriving the associated integrals and the general solution in harmonic form for both the types. We demonstrate the applicability of this method up to $2N$ coupled first order nonlinear ODEs in both the cases. Further, we  illustrate the theory with an example in each case. 
}

\maketitle
\section{Introduction}
In recent years, a great deal of interest has been shown in identifying nonlinear dynamical systems that posses isochronous property \cite{RIac}-\cite{bud}. The systems whose frequencies of oscillations are independent of the amplitudes similar to that of a linear harmonic oscillator are being termed as isochronous systems. Isochronous systems also occur in several physical situations, for example, starting from the linear harmonic oscillator to nonlinear equations like Li$\acute{e}$nard type equations. Isochronicity phenomenon has been widely studied not only for its impact in stability theory, but also for its relationship with bifurcation and boundary value problems. However, only a class of nonlinear systems exhibits isochronous oscillations. To identify and study the behaviour of such isochronous nonlinear dynamical systems, several methods have been developed in the literature, see for example Refs. \cite{cho1}-\cite{guha}. In this work, we present a novel method of generating nonlinear ordinary differential equations (ODEs) that have the characteristics of isochronous behaviour. We derive explicit general solutions and present integrals of the identified isochronous nonlinear systems. This class of isochronous nonlinear oscillators can be related to linear harmonic oscillator by introducing a nonlocal transformation in the dependent variable.

Several nonlinear dynamical systems exhibit non-isochronous oscillations. In these oscillators one can see that the frequency of oscillations depends on the amplitude. For example, the anharmonic oscillator equation $\ddot{x}+c_1x+c_2x^3=0$, with $c_1>0$, $c_2>0$, admits the solution $x(t)=Acn(\omega t,k)$, $\omega=\sqrt{c_1+c_2A^2}$, where $cn$ is the Jacobi elliptic function of modulus $k=\sqrt{c_2A^2/2(c_1+c_2A^2)}$. While deriving this solution we have considered the initial conditions $x(0)=A$, $\dot{x}(0)=0$ \cite{mlsr}. Unlike the linear harmonic oscillator equation the angular frequency $\omega$ is now not a constant but depends on the amplitude $A$. These non-isochronous dynamical systems exhibit a variety of dynamics differing from the isochronous systems. However, a class of non-isochronous dynamical systems also admit periodic oscillations of harmonic type. A notable example in this category is the Mathews-Lakshmanan nonlinear oscillator equation, $(1+\lambda x^2)\ddot{x}-\lambda x\dot{x}^2+\omega_0^2x=0$, where $\lambda$ and $\omega_0$ are constants \cite{laksh,laksh75}. This nonlinear/nonpolynomial oscillator equation has the solution $x(t)=A\cos\Omega t$, $\Omega=\omega_0/\sqrt{1+\lambda A^2}$. We have obtained this solution for the initial conditions $x(0)=A$, $\dot{x}(0)=0$ \cite{mlsr}. The angular frequency ($\Omega$) is amplitude dependent and hence the Mathews-Lakshmanan oscillator is a non-isochronous system. Because of the amplitude dependent frequency property, various responses have also been exhibited for different input conditions by the underlying dynamical system. This kind of nonisochornicity nature can be seen in various physical and biological systems. In particular, one finds that strong nonisochronicity produces various synchronization patterns as well as long irregular transient phase dynamics in networks of such oscillators in the case where the long transient dynamics has connection with the functioning of biological systems \cite{last1,last2,last3}. 

Upon observing that certain non-isochronous systems can also have periodic oscillations of harmonic type as in the case of the above Mathews-Lakshmanan oscillator (and different from that of the anharmonic oscillator mentioned earlier), we pose the following question. Is it possible to generate nonlinear isochronous and non-isochronous systems whose solutions are periodic functions of harmonic type through a single method? Our following investigation shows that one can indeed generate a class of isochronous and non-isochronous nonlinear oscillators which admits periodic solutions of harmonic type from the known integrals of the simple harmonic oscillator equation.

By investigating the time series plot of the harmonic oscillator and nonlinear isochronous oscillators we find that both of them exhibit the same frequency of oscillations but they differ only in their amplitudes. Since the angular frequency originates from the second integral of the harmonic oscillator one may consider the case where the second integral of the nonlinear systems may also be in the same form as that of the linear harmonic oscillator equation. Starting from this observation we have identified a class of nonlinear ODEs that exhibits isochronous oscillations from the second integral of the linear harmonic oscillator equation.

On the other hand, in non-isochronous systems there is an interdependency between the amplitude and the frequency of oscillations. Interestingly in the case of non-isochronous systems one can observe that the first integral has the same structure as that of the harmonic oscillator. The only difference we come across in the integrals of nonlinear systems is that the presence of an additional function in terms of the phase. This function has the variables $x$ and $\dot{x}$. By carefully examining this function we find that it is nothing but the Hamiltonian. Interestingly while deriving the general solution from the obtained integrals the same function appears as amplitude with a square root. This in turn establishes an interrelation between the amplitude and the frequency of oscillations. In this way we generate non-isochronous systems from the known integrals of the linear harmonic oscillator equation.

As far as our knowledge goes, the generalized version of the Mathews-Lakshmanan oscillator equation to $2N$ dimensions is considered for the first time. Another notable point is that the dynamics of the $2N$-coupled Mathews-Lakshmanan oscillator is studied for the non-identical system parameters (i.e. $\omega_1 \neq \omega_2$). Further, we have presented the general solution of the $2N$-coupled Mathews-Lakshmanan oscillator equation. Also we have studied the integrability of the modified Emden equation with more generalized parameters which has also not been studied previously in the literature.

We organize our work as follows. In Sec. 2, we present the method of identifying nonlinear oscillator equations which exhibit isochronous and non-isochronous properties. In Sec. 3, we consider the isochronous case and discuss the method of deriving the solution for a system of two coupled first order ODEs. We also extend the method of finding the solutions and the integrals to $2N$ coupled first-order ODEs and demonstrate it with suitable examples. In Sec. 4,  we consider the non-isochronous case and present the method of finding the solution for a system of two coupled first order ODEs. We prove the method with a physically important example, namely Mathews-Lakshmanan oscillator. Then we extend the method to $2N$-coupled first order ODEs and present the general solution for the $2N$-coupled Mathews-Lakshmanan oscillator equation. Finally, we present our conclusions in Sec. 5.
\section{Generating nonlinear systems having harmonic and anharmonic behaviour}
Let us consider a system of two coupled first order nonlinear ODE's of the form
\begin{eqnarray}
\label{nonl_gen_eqn}
\dot{x}=P(x,y),~~\dot{y}=Q(x,y),
\label{eq:1}
\end{eqnarray}
where $P$ and $Q$ are functions of $x$ and $y$ and the overdot denotes differentiation with respect to $t$. We assume that the above system of nonlinear ODEs can be connected to the linear harmonic oscillator equation, that is 
\begin{eqnarray}
\label{modeq1}
\dot{u}=\omega v,\label{mod01}~\dot{v}=-\omega u,\label{mod02}
\label{eq:2}
\end{eqnarray}
where $\omega$ is the angular frequency of oscillation. Equation (\ref{eq:2}) admits an integral of the form
\begin{equation}
I_{1l}=-\omega t+\arctan\Big(\frac{v}{u}\Big).
\label{eq:3}
\end{equation}

Suppose the nonlinear ODE (\ref{nonl_gen_eqn}) also admits the same form of integral, that is
\begin{eqnarray}
I_{1nl}&=&-\omega t+\arctan(\frac{y}{x}).
\label{eq:4}
\end{eqnarray}

Equations (\ref{eq:3}) and (\ref{eq:4}) imply that ${v}/{u}={y}/{x}$. To satisfy this equality, we correlate the variables in the form
\begin{equation}
x=\frac{u}{h(x,y)},~y=\frac{v}{h(x,y)},
\label{eq:7}
\end{equation}
where $h(x,y)$ is an arbitrary function of $x$ and $y$.

Differentiating Eq. (\ref{eq:7}) with respect to $t$ and substituting back (\ref{modeq1}) and (\ref{eq:7}) in the resultant equations and simplifying them, we obtain
\begin{eqnarray}
P(x,y)=\omega y-f(x,y)x,~~~Q(x,y)=-\omega x-f(x,y)y,
\label{eq:8}
\end{eqnarray}
where $f(x,y)=\frac{\dot{h}}{h}$.

The linear harmonic oscillator Eq.(\ref{eq:2}) has another integral (energy integral or first integral) of the form $I_{2l}=u^2+v^2$. Let the nonlinear and linear ODEs $(\ref{eq:1})$ and $(\ref{eq:2})$ have the same form of the first integral. By equating these two integrals, we find 
\begin{eqnarray}
u^2+v^2=x^2+y^2.
\label{eq:10}
\end{eqnarray}

Differentiating (\ref{eq:10}) with respect to $t$ and replacing $\dot{x}=P(x,y)$ and 
$\dot{y}=Q(x,y)$, we obtain
\begin{equation}
xP(x,y)+yQ(x,y)=0.
\label{eq:11}
\end{equation}

To fulfill the condition (\ref{eq:11}), we choose the following specific forms for $P$ and $Q$, namely
\begin{eqnarray}
P(x,y)=\omega f(x,y) y,~~~Q(x,y)=-\omega f(x,y) x,
\label{eq:12}
\end{eqnarray}
where $f$ is an arbitrary function of $x$ and $y$. We note that the form considered for $P$ and $Q$ in (\ref{eq:12}) is only a specific choice. One may consider other forms for $P$ and $Q$ also. However, in this work, we confine our attention only to this choice.

The nonlinear ODE (\ref{eq:1}) which has the right hand side of the form (\ref{eq:8}) admits isochronous behaviour whereas the nonlinear ODEs (\ref{eq:1}) which have the right hand side of the form (\ref{eq:12}) exhibit non-isochronous behaviour, as we see below in sections $3$ and $4$. 

\section{Isochronous systems}

\subsection{2-coupled isochronous systems}
\label{sec2}
To begin with, we focus our attention on the class of systems described by the form (\ref{eq:8}), that is
\begin{eqnarray}
\label{mod06}
\dot{x}=\omega y-{f(x,y)}x,~ \dot{y}=-\omega x- {f(x,y)}y.
\end{eqnarray}
From (\ref{mod06}), we can deduce the following expression, that is
\begin{eqnarray}
\label{mod08}
y\dot{x}-x\dot{y}=\omega {(x^{2}+y^{2})}.
\end{eqnarray}
Equation (\ref{mod08}) can be rewritten in the form
\begin{eqnarray}
\label{mod09}
\frac{1}{\omega}\frac{d}{dt}\bigg(\frac{x}{y}\bigg)=1+\bigg({\frac{x}{y}\bigg)^{2}}.
\end{eqnarray}
By introducing a new variable $\theta$ as
\begin{equation}
\frac{x}{y}=\tan\theta, \label{fdg}
\end{equation}
Eq. (\ref{mod09}) can be reduced to
\begin{eqnarray}
\label{mod10}
\dot \theta=\omega \Rightarrow \theta=\omega t+\delta,
\end{eqnarray}
where $\delta$ is an integration constant. From (\ref{mod10}) it is clear that for a class of ODEs considered in (\ref{mod06}), the angular velocity $(\dot\theta)$ or the angular frequency $(\omega)$ is nothing but a constant. 

Even though the angular frequency is a constant, depending on the form of $f(x,y)$, the amplitude of oscillations may decay or grow and therefore one has to identify a suitable form of $f(x,y)$ which allows periodic oscillations. After analyzing the outcome, we find that whenever the function $f(x,y)$ satisfies the condition 
\begin{eqnarray}
\label{mod10aa}
\int_0^{2\pi/\omega}f(\sin(\omega t'+\delta),\cos(\omega t'+\delta))dt'=0
\end{eqnarray}
then the underlying system (\ref{mod06}) admits an isochronous center. In other words, whenever the function $f(x,y)$ satisfies the above condition, the system (\ref{mod06}) exhibits periodic oscillations and the associated dynamical system is an isochronous one. Differing from the above, the forms which do not satisfy the condition $(\ref{mod10aa})$ display other types of oscillatory behaviour. 

\subsection{$2N$ coupled isochronous systems}
The methodology given above can be generalized to higher order ODEs also. To demonstrate this,
let us consider a system of $2N$-coupled first order nonlinear ODEs of the form [2]
	\begin{eqnarray}
	\dot{x}_{i}&=&\omega_{i}y_{i}-f(\overline{x},\overline{y})x_{i},\nonumber\\
	\label{mod43}
	\dot{y}_{i}&=&-\omega_{i}x_{i}-f(\overline{x},\overline{y})y_{i},\hspace{1cm} i=1,2,...,N,
	\end{eqnarray}
where $f(\overline{x},\overline{y})=f(x_{1},x_{2},...,x_{N},y_{1},y_{2},...,y_{N})$. We note that while extending the analysis to higher order ODEs we restrict the function $f(\bar{x},\bar{y})$ which appears in (\ref{mod43}) to be of the same form. Multiplying the first and second equations of (\ref{mod43}) by $y_{i}$ and $x_{i}$ respectively, and subtracting the resulting equations suitably, we obtain
\begin{eqnarray}
\label{mod44}
y_{i}\dot{x}_{i}-x_{i}\dot{y}_{i}=\omega_{i}(x^{2}_{i}+y^{2}_{i}), ~~~i=1,2,...,N.
\end{eqnarray}
Dividing Eqs. (\ref{mod44}) by $y^{2}_{i}$, $i=1,..., N,$ and rewriting it, we find
\begin{eqnarray}
\label{mod45}
\frac{d}{dt}\Big(\frac{x_{i}}{\omega_{i}y_{i}}\Big)=1+\Big(\frac{x_{i}}{y_{i}}\Big)^{2},\hspace{1cm} i=1,2,...,N.
\end{eqnarray}
Upon introducing the angle variable $\displaystyle \theta_{i}=\tan^{-1}\Big(\frac{x_{i}}{y_{i}}\Big)$, Eq. (\ref{mod45}) becomes
\begin{eqnarray}
\label{mod46}
\dot\theta_{i}=\omega_{i},\hspace{1cm} i=1,2,...,N.
\end{eqnarray}
Equation (\ref{mod46}) reveals that the angular frequency is independent of the amplitudes of oscillations, irrespective of the form of $f(\overline{x},\overline{y})$.

In the following, we demonstrate the existence of $2N$ independent integrals for the system of equations (\ref{mod43}). To construct the integrals of Eq. (\ref{mod43}), we recall 
\begin{eqnarray}
\label{mod47}
\frac{y_{i}}{x_{i}}=\cot(\omega_{i}t+\delta_{i}),\hspace{1cm}i=1,2,...,N.
\end{eqnarray}
From Eq. (\ref{mod47}), we can identify $N$ time dependent integrals which are of the form
\begin{eqnarray}
\label{mod48}
\delta_{i}=\cot^{-1}\bigg[\frac{y_{i}}{x_{i}}\bigg]-\omega_{i}t,\hspace{1cm}i=1,2,...,N.
\end{eqnarray}
To determine the remaining $N$ integrals of Eq. (\ref{mod43}), we proceed as follows. 
From Eq.(\ref{mod43}), we can identify
\begin{eqnarray}
\label{mod48a}
\frac{x_j\dot{x}_j+y_j\dot{y}_j }{x_j^2+y_j^2}=\frac{x_N \dot{x}_N+y \dot{y}_N }{x_N^2+y_N^2}.\hspace{1cm}j=1,2,...,N-1.
\end{eqnarray}
Rewriting Eq. (\ref{mod48a}) as $ \frac{d}{dt} \left[ \log(x_j^2 + y_j^2) / 2\right]=\frac{d}{dt} \left[ \log(x_N^2 + y_N^2) / 2\right] $ and integrating and rewriting the resultant integrals for $j=1,2,\cdots,N-1$, we obtain $N-1$ time independent integrals of the form
\begin{eqnarray}
\label{mod48b}
I_j=\frac{x_j^2+y_j^2}{x_N^2+y_N^2}.\hspace{1cm}j=1,2,...,N-1.
\end{eqnarray}
By utilizing the obtained $(N-1)$ integrals we can reduce the set of $2N$ first order equations to a system of two coupled first order ODEs in the variables $x_N$ and $y_N$. In the following, we demonstrate this. \\

From Eq.(\ref{mod47}), we find 
\begin{equation}
\frac{x_{i}}{y_{i}}=\tan(\theta_{i}),\; \theta_{i}=\omega_{i}t+\delta_{i}.
\end{equation}

Using the well-known trigonometric identities, that is  $\sin{\theta_{i}}=\frac{\tan{\theta_{i}}}{\sqrt{1+\tan^{2}{\theta_{i}}}}$ and $\cos{\theta_{i}}=\frac{\cot{\theta_{i}}}{\sqrt{1+\cot^{2}{\theta_{i}}}}$, we can identify
	\begin{eqnarray}
	\label{mod49}
	&&\sin(\omega_{i}t+\delta_{i})=\frac{x_{i}}{\sqrt{x^{2}_{i}+y^{2}_{i}}},\\
	\label{mod50}
	&&\cos(\omega_{i}t+\delta_{i})=\frac{y_{i}}{\sqrt{x^{2}_{i}+y^{2}_{i}}},\hspace{1cm}i=1,2,...,N.
	\end{eqnarray}
From (\ref{mod49}), we find 
\begin{eqnarray}
\label{mod50b}
x_i(t)&=&\sin(\omega_{i}t+\delta_{i})(\sqrt{x^{2}_{i}+y^{2}_{i}})\nonumber\\
&=&\sqrt{I_i}\sin(\omega_{i}t+\delta_{i})(\sqrt{x^{2}_{N}+y^{2}_{N}}) \nonumber\\
&=&\sqrt{I_i}\frac{\sin(\omega_{i}t+\delta_{i})}{\sin(\omega_{N}t+\delta_{N})} x_N.
\end{eqnarray}
Similarly
\begin{eqnarray}
\label{mod50c}
y_i(t)&=&\sqrt{I_i}\frac{\cos(\omega_{i}t+\delta_{i})}{\cos(\omega_{N}t+\delta_{N})} y_N.
\end{eqnarray}
With the help of Eqs. (\ref{mod50b}) and (\ref{mod50c}) one can replace the variables $x_i$ and $y_i$, $i=1,2,\dots N-1$, in terms of $t,x_N$ and $y_N$. As a result, we find 
\begin{eqnarray}
\label{mod50d}
f(\overline{x},\overline{y})=f(x_1,x_2,\dots x_N,y_1,y_2,\dots y_N)=f(t,x_N,y_N).
\end{eqnarray}
Now confining our attention on the last two ODEs in $(\ref{mod43})$, we find
	\begin{eqnarray}
	\label{mod42aa}
	\dot{x}_{N}&=&\omega_{N}y_{N}-f(t,x_N,y_N)x_{N},\\
	\label{mod43aa}
	\dot{y}_{N}&=&-\omega_{N}x_{N}-f(t,x_N,y_N)y_{N}.
	\end{eqnarray}
The above equation cannot be integrated explicitly for an arbitrary form of $f(t,x_N,y_N)$. For particular forms of $f(t,x_N,y_N)$, we can solve the above equation. The theory developed above helps to identify a class of nonlinear isochronous systems and their general solution.

\subsection{Examples}

In the following, with appropriate choices of the function $f(x, y)$, we demonstrate through specific examples how specific nonlinear integrable dynamical systems corresponding to $N = 1, 2$ and arbitrary integer can be identified which admit isochronous oscillations.  The crucial property which we impose on the choice of the function $f(x, y)$ is that it should satisfy the integral condition given by Eq.(\ref{mod10aa}). The condition will also hold good when $N$ is arbitrary, as for example in Eqs.(\ref{mod42aa}) and (\ref{mod43aa}), the condition (\ref{mod10aa}) will be imposed on the form of $f(t, x_N, y_N)$, as illustrated in the examples given below.

\subsubsection{Example 1}

To start with we consider the simplest case, namely $N =2$. Based on our general prescription given above, we consider the choice 
\begin{eqnarray}
\label{mod12}
{f(x,y)}=a_{1}x^{q}+a_{2}y^{q},
\end{eqnarray}
where $q$ is a positive integer and  $a_{1}$ and $a_{2}$ are constants.
Substituting the expression $x=y\tan\theta$, with $\theta=\omega t+\delta$, the second expression in Eq. (\ref{mod06}) can be brought to the form
\begin{eqnarray}
\label{mod13}
\dot{y}={(a_{1}\tan^{q}{(\omega t+\delta)}+a_{2})y^{q+1}}-\omega \tan{(\omega t+\delta)}y.
\end{eqnarray}
  
Equation (\ref{mod13}) is of Bernoulli form. The general solution of Eq. (\ref{mod13}) reads
\begin{eqnarray}
\label{mod14}
y(t)=\frac{\cos{(\omega t+\delta)}}{{\Big(I_{1}-q\int \big(a_{1}{\sin}^{q}{(\omega t+\delta)}+a_{2}{\cos}^{q}{(\omega t+\delta)\big)dt\Big)^{\frac{1}{q}}}}}
\label{18}
\end{eqnarray}

from which the expression for $x(t)$ can be deduced as
\begin{eqnarray}
\label{mod15}
x(t)=\frac{\sin{(\omega t+\delta)}}{{\Big(I_{1}-q\int \big(a_{1}{\sin}^{q}{(\omega t+\delta)}+a_{2}{\cos}^{q}{(\omega t+\delta)\big)dt\Big)^{\frac{1}{q}}}}}.
\label{19}
\end{eqnarray}

The integral which is appearing in the denominator in expressions Eqs. (\ref{18}) and (\ref{19}) can be evaluated depending on whether $q$ is an odd or even integer. 

 We note here that  the function given in (\ref{mod12}) satisfies the condition (\ref{mod10aa}) only for odd values of $q$ so that the solution exhibits periodic  behaviour. For even value of $q$ the function (\ref{mod12}) does not satisfy the condition (\ref{mod10aa}) and the solution exhibits damped behaviour. In the following, we present the solutions for both the cases.

\subsubsection{Case 1: $q$ is an odd integer $(q=2m+1)$}
\label{3}
For this choice, we obtain the solution $x(t)$ and $y(t)$ in the form \cite{tablebook}
\begin{eqnarray}
\label{mod20}
x(t)=\sin{(\omega t+\delta)}{\Bigg[I_{1}-q(b_1(t)+b_2(t))\Bigg]}^\frac{-1}{2m+1},\nonumber\\
\label{mod21}
y(t)=\cos{(\omega t+\delta)}{\Bigg[I_{1}-q(b_1(t)+b_2(t))\Bigg]}^\frac{-1}{2m+1},
\end{eqnarray}
with 
\begin{eqnarray}
b_1(t)= \frac{a_1}{\omega} \bigg[\frac{-\cos\tau}{2m+1}\bigg(\sin^{2m}\tau+\sum_{k=0}^{m-1}s_k^1\sin^{2m-2k-2}\tau\bigg)\bigg], \nonumber\\
b_2(t)=\frac{a_2}{\omega} \bigg[\frac{\sin\tau}{2m+1}\bigg(\cos^{2m}\tau+\sum_{k=0}^{m-1}s_k^1\cos^{2m-2k-2}\tau\bigg)\bigg].
\label{eq:sol2aa}
\end{eqnarray}
Here $\tau=\omega t+\delta $ and $s_k^1=(2^{k+1}m(m-1)\ldots (m-k))/((2m-1)(2m-3)\ldots (2m-2k-1))$.

\subsubsection{Case 2: $q$ is even integer $(q=2m)$}
In this case, the solution (\ref{mod13}) takes the form \cite{tablebook}
\begin{eqnarray}
\label{mod28}
x(t)=\sin{(\omega t+\delta)}{\Bigg[I_{1}-q(b_3(t)+b_4(t))\Bigg]}^\frac{-1}{2m},\nonumber\\
\label{mod29}
y(t)=\cos{(\omega t+\delta)}{\Bigg[I_{1}-q(b_3(t)+b_4(t))\Bigg]}^\frac{1}{2m},
\end{eqnarray}
where 
\begin{eqnarray}
b_3(t)= \frac{a_1}{\omega} \bigg[\frac{-\cos\tau}{2m}\bigg(\sin^{2m-1}\tau+\sum_{k=0}^{m-1}s_k^2\sin^{2m-2k-1}\tau\bigg)+\frac{(2m-1)!!}{2^mm!}\tau\bigg],\nonumber\\
b_4(t)= \frac{a_2}{\omega} \bigg[\frac{\sin\tau}{2m}\bigg(\cos^{2m-1}\tau+\sum_{k=0}^{m-1}s_k^2\cos^{2m-2k-1}\tau\bigg)+\frac{(2m-1)!!}{2^mm!}\tau\bigg].
\label{eq:sol2ab}
\end{eqnarray}
Here  $s_k^2=((2m-1)(2m-3)\ldots (2m-2k+1))/(2^{k}(m-1)(m-2)\ldots (m-k))$. 

For odd and even values of $q$ the system (\ref{mod06}) with the form of $f(x,y)$ given in (\ref{mod12}) exhibits periodic and damped oscillations, respectively. The solution plot given in Figure 1 confirms the same.\\
\begin{figure*}	
\centering
\includegraphics{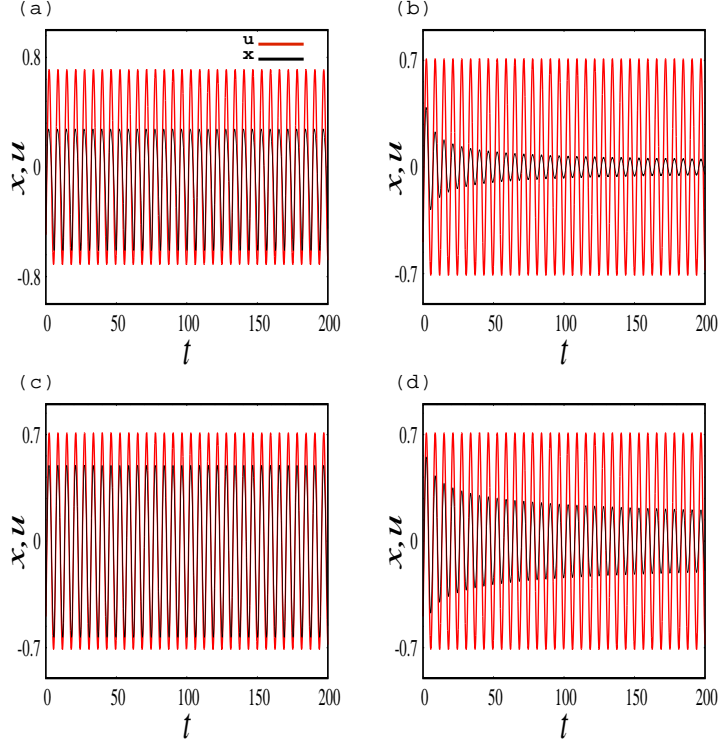}
\caption{Time series plot of equation (\ref{mod06}) (black line) for (a) $q=1$, (b) $q=2$, (c) $q=3$, and (d) $q=4$ show both periodic (for $q$ odd values) and damped (for $q$ even values) oscillation. Red/grey line is the time series plot of equation (\ref{modeq1}). Initial conditions and parameter values are $x(0)=u(0)=-0.5$, $y(0)=v(0)=0.5$ and $a=b= \omega=1.0$.}  
\end{figure*}

{\it Case: $f(x,y)=ax+by,~q=1$}\\

For the choice $f(x,y)=ax+by$ ($a_{1}=a$, $a_{2}=b$ and $q=1$ in Eq. (\ref{mod12})), the coupled first order nonlinear ODEs (\ref{eq:1}) read as
\begin{eqnarray}
\label{mod30_n}
\dot{x}=\omega y-(ax+by)x,\label{mod31_n}~~~\dot{y}=-\omega x-(ax+by)y.
\end{eqnarray}
\label{1mod32_nn}
The above system is a special case of the most general quadratic system, that was studied by Loud \cite{loud}.

Rewriting Eq. (\ref{mod30_n}) in second order form, we find
\begin{eqnarray}
\label{mod33_n}
\ddot{x}+3ax\dot{x}+a^2x^{3}+\omega x(\omega-bx)+\frac{1}{(\omega-bx)}(2b\dot{x}^2+3abx^2\dot{x}+a^2bx^4)=0.
\end{eqnarray}
To the best of our knowledge, the general solution of the above generalized version of the modified Emden equation is reported here for the first time. The general solution of Eq. (\ref{mod33_n}) is given by (vide Eqs.(\ref{mod14}) and (\ref{mod15}))
\begin{eqnarray}
x(t)&=&\frac{\omega \sin(\omega t+\delta)}{\omega I_1+a \cos (\omega t+\delta)-b \sin(\omega t+\delta)},\nonumber
\label{eq:20}\\
y(t)&=&\frac{\omega \cos(\omega t+\delta)}{\omega I_1+a \cos (\omega t+\delta)-b \sin(\omega t+\delta)},~~\mathrm{I_1:constant}.
\end{eqnarray}

By substituting $f(x,y)=x$ in (\ref{eq:8}), we find $\dot{x}=\omega y-x^2$, $\dot{y}=-\omega x-xy$. Rewriting this as a single second order ODE in the variable $x$, we obtain the modified Emden equation \cite{annapre}, that is $\ddot{x}+3x\dot{x}+x^3+\omega^2x=0$. The solution of this equation can be extracted from the Eqs. $(\ref{18})$ and (\ref{mod15}) by fixing $q=1$ and $a_2=0$. The resultant form agrees with the one reported in the literature \cite{annapre}.

For the choice $q=2$ in Eq. (\ref{mod12}), the resultant system turns out to be a cubic system. The isochronous cases of these systems have already been studied in the literature, see for example Refs. \cite{chav,chou2}. 
\begin{figure*}
\centering
\resizebox{0.7\textwidth}{!}{
\includegraphics{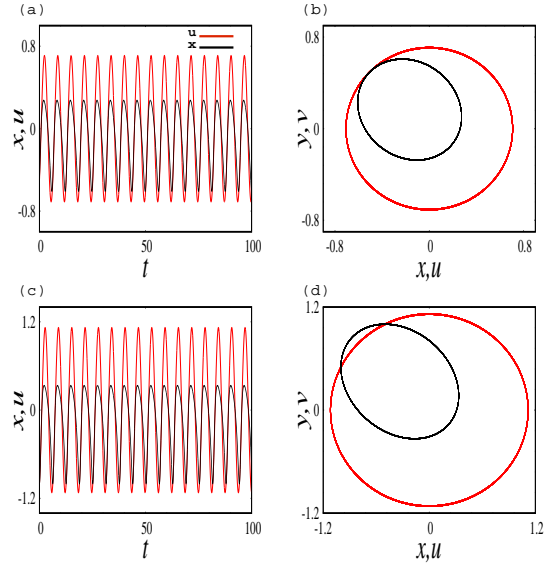}}
\caption{Time series plot of equations (\ref{modeq1})  and (\ref{mod30_n}) (red and black, respectively) exhibiting periodic oscillations with the same frequency with different amplitudes for the initial conditions (a) $x(0)=u(0)=-0.5$ and $y(0)=v(0)=0.5$ and (b) $x(0)=u(0)=-1$ and $y(0)=v(0)=0.5$. Corresponding phase space portraits are given in (b) and (d) for $a=b= \omega=1.0$.}  
	\end{figure*}

Figure 2 shows the periodic oscillations admitted by Eqs. (\ref{modeq1})  and (\ref{mod30_n}) for two different sets of initial conditions. One may note that the frequency of oscillations of (\ref{mod33_n}) is independent of the initial conditions as in the case of the linear harmonic oscillator. However, the amplitude of oscillations of the nonlinear system differs from the linear one.

\subsubsection{Example 2}
 For the case $n=2$, the equation of motion can be written in the form 
	    \begin{eqnarray}
    \dot{x}_1&=&\omega_1 y_1-f(\bar{x},\bar{y})x_1,\nonumber\\
    \dot{x}_2&=&\omega_2 y_2-f(\bar{x},\bar{y})x_2,\nonumber\\
    \dot{y}_1&=&-\omega_1 x_1-f(\bar{x},\bar{y})y_1,\nonumber\\
    \dot{y}_2&=&-\omega_2 x_2-f(\bar{x},\bar{y})y_2,\label{mod432}
    \end{eqnarray}
  	where $f(\bar{x},\bar{y})=f(x_1,x_2,y_1,y_2)$. To proceed further, we assume a specific form for the function $f(x,y)$, namely
\begin{equation}
f=a_1x_1^q+b_1y_1^q+a_2x_2^q+b_2y_2^q,
\label{40}
\end{equation}
where $q$ is an integer and $a_i$'s, $b_i$'s, $i=1,2$, are constants. Now the equation of motion for $q=1$ turns out to be
    \begin{eqnarray}
	\dot{x}_1&=&w_1 y_1 - (a_1 x_1 + b_1 y_1 + a_2 x_2 + b_2 y_2)x_1,\nonumber\\
	\dot{y}_1&=&-w_1 x_1 - (a_1 x_1 + b_1 y_1 + a_2 x_2 + b_2 y_2) y_1,\nonumber\\
	\dot{x}_2&=&w_2 y_2 - (a_1 x_1 + b_1 y_1 + a_2 x_2 + b_2 y_2) x_2,\nonumber\\
	\dot{y}_2&=&-w_2 x_2 - (a_1 x_1 + b_1 y_1 + a_2 x_2 + b_2 y_2) y_2,
\label{52}
	\end{eqnarray}
where $w_i$ and $a_j$'s, $b_j$'s $i=1,2$ and $j=1,2,3,4$, are all real arbitrary constants.
	Following the procedure discussed in the previous section with $q=1$, we can identify the general solution of $(\ref{52})$ in the form
		\begin{eqnarray}
		&&x_1(t)=\frac{\sqrt{I_1} w_1 w_2 \sin \theta_1}{G_3},~~
		y_1(t)=\frac{\sqrt{I_1} w_1 w_2 \cos \theta_1}{G_3},\\
		&&x_2(t)=\frac{ w_1 w_2 \sin \theta_2}{G_3},~~
		y_2(t)=\frac{ w_1 w_2 \cos \theta_2}{G_3},
		\label{eq:}
		\end{eqnarray}
	where $\theta_1=w_1 t+\delta_1$, $\theta_2=w_2 t+\delta_2$, $G_3=I_2 w_1 w_2-a_1\sqrt{I_1}w_2\cos \theta_1-a_2w_1\cos \theta_2+b_1\sqrt{I_1}w_2\sin \theta_1+b_2w_1\sin \theta_2$.\\
	
\par For the choice, $q=2$, we consider an equation which is of the form
		\begin{eqnarray}
		&&\dot{x}_1=w_1 y_1 - (a_1 x_1^2 + b_1 y_1^2 + a_2 x_2^2 + b_2 y_2^2)x_1,\nonumber\\
		&&\dot{y}_1=-w_1 x_1 - (a_1 x_1^2 + b_1 y_1^2 + a_2 x_2^2 + b_2 y_2^2) y_1,\nonumber\\
		&&\dot{x}_2=w_2 y_2 - (a_1 x_1^2 + b_1 y_1^2 + a_2 x_2^2 + b_2 y_2^2) x_2,\nonumber\\
		&&\dot{y}_2=-w_2 x_2 - (a_1 x_1^2 + b_1 y_1^2 + a_2 x_2^2 + b_2 y_2^2) y_2,
		\label{neweq123}
		\end{eqnarray}
where $w_i$ and $a_j$'s, $b_j$'s $i=1,2$ and $j=1,2,3,4$, are all real arbitrary constants. The general solution of $(\ref{neweq123})$ can be identified in a straightforward manner as
	\begin{eqnarray}
	&&x_1(t)=\frac{\sqrt{2 I_1 w_1 w_2} \sin \theta_1}{G_4},~~
	y_1(t)=\frac{\sqrt{2 I_1 w_1 w_2} \cos \theta_1}{G_4},\nonumber\\
	&&x_2(t)=\frac{\sqrt{2  w_1 w_2} \sin \theta_2}{G_4},~~
	y_2(t)=\frac{\sqrt{2  w_1 w_2} \cos \theta_2}{G_4},
	\end{eqnarray}
	where $\theta_1=w_1 t+\delta_1$, $\theta_2=w_2 t+\delta_2$,
 $G_4=\big[2 w_1 w_2 I_2+2((a_1+b_1)I_1w_2\theta_1+(a_2+b_2) w_1\theta_2)-(a_1-b_1)I_1w_2\sin 2\theta_1+(b_2-a_2)w_1\sin 2\theta_2\big]^{\frac{1}{2}}$, $I_1$ and $I_2$ are the integration constants. The above solution leads to damped type solution which is similar to the Figure 1 for the values $q=2$ and $q=4$.

\subsubsection{Example 3} 
For $2N$ coupled system, we consider a specific form for $f$ as
\begin{eqnarray}
\label{mod12ab}
{f(\overline{x},\overline{y})}=\sum_{j=1}^N a_{j}x_j^{q}+b_{j}y_j^{q},
\end{eqnarray}
where $a_j$'s and $b_j$'s are constants.
The equation of motion becomes
\begin{eqnarray}
\dot{x}_i&=&\omega_{i}y_{i}-\big[\sum_{j=1}^N a_{j}x_j^{q}+b_{j}y_j^{q}\big]x_{i},\nonumber\\
\dot{y}_{i}&=&-\omega_{i}x_{i}-\big[ \sum_{j=1}^N a_{j}x_j^{q}+b_{j}y_j^{q}\big]y_{i},\hspace{1cm} i=1,2,...,N,
\label{neqn_iso}
\end{eqnarray}
 For this form of $f$, we have to find $x_N$ and $y_N$ to write the general solution of (\ref{neqn_iso}) (vide Eqs.(\ref{mod50b}) and (\ref{mod50c})). Eq.(\ref{mod43aa}) becomes
\begin{eqnarray}
\label{mod43ab}
\dot{y}_{N}(t)&=&-\omega_{N}\tan(\omega_{N}t+\delta_{N})y_{N} -\sum_{i=1}^N \bigg(a_j I_j^{\frac{q}{2}}\frac{\sin^q(\omega_{i}t+\delta_{i})}{\cos^q(\omega_{N}t+\delta_{N})} \nonumber\\ &&+b_j I_j^{\frac{q}{2}}\frac{\cos^q(\omega_{i}t+\delta_{i})}{\cos^q(\omega_{N}t+\delta_{N})}\bigg) y_{N}^{q+1}.
\end{eqnarray}
Equation $(\ref{mod43ab})$ is of Bernoulli type and the general solution is given by 
\begin{eqnarray}
y_N(t)&=&\frac{\cos(\omega_N t+\delta_N)}{\bigg[I_N-q\int \sum_{j=1}^N (a_j I_j^{\frac{q}{2}}\sin^q(\omega_{i}t+\delta_{i})
	+b_j I_j^{\frac{q}{2}}\cos^q(\omega_{i}t+\delta_{i}))dt\bigg]^{\frac{1}{q}}}. 
\label{eq:sol1}
\end{eqnarray}
Substituting this back in $x_{N}=y_N\tan(\omega_{N}+\delta_{N})$, we find the following expression for $x_N(t)$, that is
\begin{eqnarray}
	x_N(t)&=&\frac{\sin(\omega_N t+\delta_N)}{\bigg[I_N-q\int \sum_{j=1}^N (a_j I_j^{\frac{q}{2}}\sin^q(\omega_{i}t+\delta_{i})
		+b_j I_j^{\frac{q}{2}}\cos^q(\omega_{i}t+\delta_{i}))dt\bigg]^{\frac{1}{q}}}. 
\end{eqnarray}
Here also, the integral which is appearing in the denominator can be evaluated depending on whether $q$ is an odd or even integer. By substituting the above form of $x_N$ and $y_N$ in Eqs. (\ref{mod50b}) and (\ref{mod50c}), we get the general solution of Eq. (\ref{neqn_iso}). Now the general solution turns out to be 
\begin{eqnarray}
x_i&=&\frac{\sqrt{I_i}\sin(\omega_{i}t+\delta_{i})}{\bigg[I_N-q\int \sum_{j=1}^N (a_j I_j^{\frac{q}{2}}\sin^q(\omega_{i}t+\delta_{i})
		+b_j I_j^{\frac{q}{2}}\cos^q(\omega_{i}t+\delta_{i}))dt\bigg]^{\frac{1}{q}}},\nonumber\\
y_i&=&\frac{\sqrt{I_i}\cos(\omega_{i}t+\delta_{i})}{\bigg[I_N-q\int \sum_{j=1}^N (a_j I_j^{\frac{q}{2}}\sin^q(\omega_{i}t+\delta_{i})
		+b_j I_j^{\frac{q}{2}}\cos^q(\omega_{i}t+\delta_{i}))dt\bigg]^{\frac{1}{q}}}.
\end{eqnarray}

For odd integers ($q=2m+1$), the integral appearing in the denominator can be integrated to yield
\begin{eqnarray}
\int \sum_{j=1}^N (a_j I_j^{\frac{2m+1}{2}}\sin^{2m+1}(\omega_{i}t+\delta_{i})
+b_j I_j^{\frac{2m+1}{2}}\cos^{2m+1}(\omega_{i}t+\delta_{i}))dt\nonumber\\=\sum_{j=1}^N \frac{a_j}{\omega_j} I_j^{\frac{2m+1}{2}}\bigg[\frac{-\cos\tau_j}{2m+1}\bigg(\sin^{2m}\tau_j+\sum_{k=0}^{m-1}s_k^1\sin^{2m-2k-2}\tau_j\bigg)\bigg]\nonumber\\
+\sum_{j=1}^N \frac{b_j}{\omega_j} I_j^{\frac{2m+1}{2}}\bigg[\frac{\sin\tau_j}{2m+1}\bigg(\cos^{2m}\tau_j+\sum_{k=0}^{m-1}s_k^1\cos^{2m-2k-2}\tau_j\bigg)\bigg],
\label{eq:sol2}
\end{eqnarray}
where $\tau_j=\omega_{j}t+\delta_{j}$ and $s_k^1=(2^{k+1}m(m-1)\ldots (m-k))/((2m-1)(2m-3)\ldots (2m-2k-1))$.

For even integers ($q=2m$), we find
\begin{eqnarray}
\int \sum_{j=1}^N (a_j I_j^{m}\sin^{2m}(\omega_{i}t+\delta_{i})
+b_j I_j^{m}\cos^{2m}(\omega_{i}t+\delta_{i}))dt\nonumber\\=\sum_{j=1}^N \frac{a_j}{\omega_j} I_j^{\frac{q}{2}}\bigg[\frac{-\cos\tau_j}{2m}\bigg(\sin^{2m-1}\tau_j+\sum_{k=0}^{m-1}s_k^2\sin^{2m-2k-1}\tau_j\bigg)+\frac{(2m-1)!!}{2^mm!}\tau_j\bigg]\nonumber\\
+\sum_{j=1}^N \frac{b_j}{\omega_j} I_j^{\frac{q}{2}}\bigg[\frac{\sin\tau_j}{2m}\bigg(\cos^{2m-1}\tau_j+\sum_{k=0}^{m-1}s_k^2\cos^{2m-2k-1}\tau_j\bigg)+\frac{(2m-1)!!}{2^mm!}\tau_j\bigg],\nonumber\\
\label{eq:sol2}
\end{eqnarray}
where  $s_k^2=((2m-1)(2m-3)\ldots (2m-2k+1))/(2^{k}(m-1)(m-2)\ldots (m-k))$.  The above expressions can be derived by considering the reduction formulas for the integrals of $\sin^{n}\theta$ and $\cos^{n}\theta$ and they can be written as
\begin{eqnarray}
\int{\sin^{n}\theta d\theta}&=&-\frac{1}{n}\sin^{n-1}\theta \cos \theta+\frac{n-1}{n}\int{\sin^{n-2}\theta d\theta}\nonumber,\\
\int{\cos^{n}\theta d\theta}&=&\frac{1}{n}\cos^{n-1}\theta \cos \theta+\frac{n-1}{n}\int{\cos^{n-2}\theta d\theta}.\nonumber
\label{eq:}
\end{eqnarray}

\section{Non-isochronous Systems}
	  In this section, we present the method of generating nonlinear oscillators having non-isochronous property. Differing from isochronous systems, these systems have the property of amplitude dependent frequency of oscillations. We point out below that this property essentially arises from the fact that the the form of the function $f(x,y)$ should be related to an integral of motion and we make the simplest of choice $f(x,y) = f(I)$, where $I =x^2+y^2$, the energy integral of the linear harmonic oscillator. the details are given in the following. To make our investigations systematic, we start our discussion by considering a system of two coupled first order ODEs.
	  
		\subsection{2-Coupled Non-iscochronous Systems}
	\label{sec1}
	In this case the nonlinear ODEs take the form (see Eq.(\ref{eq:12})): 
	\begin{eqnarray}
	\dot{x}=\omega f(x,y) y,\label{mod59}~~	\dot{y}=-\omega f(x,y) x.\label{mod60}
	\end{eqnarray}
	\label{1mod60}
	From Eq. (\ref{mod59}) we find
	\begin{eqnarray}
	\label{mod62}
	\frac{y\dot{x}-x\dot{y}}{(x^{2}+y^{2})}=f(x,y) \omega.
	\end{eqnarray}
     Upon introducing the angle variable expression (\ref{fdg}) in Eq. (\ref{mod62}), the latter equation becomes
	\begin{eqnarray}
	\label{mod66}
	\dot\theta=\Omega(x,y), 
	\end{eqnarray}
  where $\Omega=\omega f(x,y)$.
	One may observe that unlike the isochronous case, here the angular velocity is not a constant but a function of $x$ and $y$. This constraint introduces a connection between amplitude and the frequency, as we see below.
		
	Now we analyze the structure of the form $f(x,y)$. Since it appears inside the phase it should be a constant.  So we can consider any form of a constant associated with the function $f(x,y)$ and consequently the resultant system becomes a nonisochronous one. Obviously, the simplest choice one can make for the function $f(x,y)$ is that it can be a function of an integral of motion, that is $f(x,y)=f(I)$. This considered integral should be independent of $t$. Since we are studying a second order system the obvious choice is the first integral ($I=H=x^2+y^2$). With these arguments we choose $f(x,y)=f(H(x,y))$.


 	To obtain the general solution of (\ref{mod60}), we follow the same procedure as discussed in the case of isochronous systems. Considering $H=x^{2}+y^{2}$ and using the expression (\ref{fdg}), we find
	\begin{eqnarray}
	\label{mod69}
	y^{2}=H-y^{2}\tan^{2}(\Omega t+\delta),
	\end{eqnarray}
	from which we can fix
	\begin{eqnarray}
	\label{mod70}
	y=\sqrt{H}\cos(\Omega t+\delta).
	\end{eqnarray}
	Now substituting (\ref{mod70}) in (\ref{fdg}), we obtain
	\begin{eqnarray}
	\label{mod71}
	x=\sqrt{H}\sin(\Omega t+\delta).
	\end{eqnarray}
	Replacing $\Omega=\omega f(H)$ in Eqs. (\ref{mod70}) and (\ref{mod71}) we end up with
	\begin{eqnarray}
		\label{eqnneq4}
	\label{mod72}
	x(t)=\sqrt{H}\sin\big(f(H)\omega t+\delta\big),~~~
	\label{mod73}
	y(t)=\sqrt{H}\cos\big(f(H)\omega t+\delta\big).
	\end{eqnarray}
The methodology given above clearly demonstrates the interdependency between amplitude and the frequency of oscillations in a class of nonlinear non-isochronous systems.

\subsection{Generalization to $2N$ coupled first order non-isochronous nonlinear equations}
In this section, we generalize the results given in the previous subsection to the system of $N$ coupled nonlinear oscillators of Li$\acute{e}$nard type. Here we consider the equation of motion in the form
	\begin{eqnarray}
	\label{mod81}
	\dot{x}_{i}=f(\overline{x},\overline{y})\omega_{i}y_{i},~~
	\dot{y}_{i}=-f(\overline{x},\overline{y})\omega_{i}x_{i},\;\;\;\;i=1,2,\cdots,N,
	\end{eqnarray}
where $f(\overline{x},\overline{y})=f(x_{1},x_{2},...,x_{N},y_{1},y_{2},...,y_{N})$. To construct the general solution of Eq. (\ref{mod81}), we proceed in the following way. 

From Eq. (\ref{mod81}) we can derive the following expressions, namely
\begin{eqnarray}
\label{mod83}
y_{i}\dot{x}_{i}-x_{i}\dot{y}_{i}=\omega_{i}f(\overline{x},\overline{y})(x_{i}^{2}+y_{i}^{2}),\hspace{1cm} i=1,2,...,N.
\end{eqnarray}
Let us choose $f(\overline{x},\overline{y})=f(\overline{H})=f(H_1,H_2,\cdots,H_N)$, where
\begin{eqnarray}
\label{mod84}
x_{i}^{2}+y_{i}^{2}=H_{i}, \qquad i=1,2,\dots 
\end{eqnarray}
Substituting the form $f(\overline{x},\overline{y})=f(\overline{H})$ in (\ref{mod83}), we obtain
\begin{eqnarray}
\label{mod85}
y_{i}\dot{x}_{i}-x_{i}\dot{y}_{i}=\omega_{i}f(\overline{H})((x_{i}^{2}+y_{i}^{2}),\hspace{1cm} i=1,2,...,N.
\end{eqnarray}
Following the steps given in Sec. 5, we derive the solution of Eq. (\ref{mod81}) in the form
	\begin{eqnarray}
	\label{mod86}
	x_{i}(t)=\sqrt{H_{i}}\sin\Bigg(f(\overline{H})\omega_{i}t+\delta_{i}\Bigg),\hspace{1cm}i=1,2,...,N,\\
	\label{mod87}
	y_{i}(t)=\sqrt{H_{i}}\cos\Bigg(f(\overline{H})\omega_{i}t+\delta_{i}\Bigg),\hspace{1cm}i=1,2,...,N,
	\end{eqnarray}
where $\delta_{i},~ i=1,2,3,...,2N$, are integration constants. Out of the $2N$ integrals, $N$ integrals show explicit time dependency as
\begin{eqnarray}
\label{mod88}
\delta_{i}=\cot^{-1}\bigg[\frac{y_{i}}{x_{i}}\bigg]-\Omega_{i}t=\cot^{-1}\bigg[\frac{y_{i}}{x_{i}}\bigg]-f(\overline{H})\omega_{i}t,\hspace{1cm}i=1,2,...,N
\label{mod89}
\end{eqnarray}
and the remaining $N$ integrals exhibit time independency in the form
\begin{eqnarray}
\label{mod90}
H_{i}=x_{i}^{2}+y_{i}^{2},\hspace{1cm}i=1,2,...,N.
\end{eqnarray}
These $2N$ functionally independent integrals provide the general solution of (\ref{mod81}).
\subsection{Examples: Nonisochronous oscillators}
\subsubsection{Example: Mathews-Lakshmanan Oscillator}
To demonstrate the theory presented above, we choose $f(x,y)=[1+\lambda (x^2+y^2)]^{-\frac{1}{2}}$ so that Eq. (\ref{mod59}) becomes
	\begin{eqnarray}
	\dot{x}=\frac{\omega y}{\sqrt{1+\lambda(x^{2}+y^{2})}},~~~
	\label{mod94}
	\dot{y}=-\frac{\omega x}{\sqrt{1+\lambda(x^{2}+y^{2})}}.
	\end{eqnarray}
where $\lambda$ and $\omega^2$ are arbitrary constants. Rewriting Eq. (\ref{mod94}) as a second order differential equation in $x$, we obtain the celebrated Mathews-Lakshmanan oscillator equation \cite{laksh}
\begin{eqnarray}
\label{mod96}
(1+ \lambda x^{2})\ddot{x}- \lambda x \dot{x}^{2}+\omega^{2}x=0.
\end{eqnarray}

 The restriction $\lambda =0$ gives harmonic oscillator equation. Eq.(\ref{mod96}) admits the Lagrangian and Hamiltonian structures, respectively, of the form 
\begin{eqnarray}
L=\frac{ \dot{x}^2-\omega^{2} x^2}{2(1+ \lambda x^{2})}~ \mathrm{and}~ H=\frac{1}{2}[p^2(1+\lambda x^2)+\frac{\omega^2x^2}{(1+\lambda x^2)}],
\end{eqnarray} 
where the canonically conjugate momentum $p$ is given by $p=\frac{\dot{x}}{(1+\lambda x^2)}$.

During the past few years several studies have been made to explore various physical and mathematical properties of the nonlinear oscillator equation (\ref{mod96}), see for example Refs.\cite{Venk}-\cite{newone}. 

The general solution of (\ref{mod94}) can be identified from Eq. (\ref{eqnneq4}) in the form
		\begin{eqnarray}
	x(t)&=&A\sin\big((1+\lambda A^2)^{-1/2}\omega t+\delta\big),\nonumber\\
	y(t)&=&A\cos\big((1+\lambda A^2)^{-1/2}\omega t+\delta\big),\label{ml_soln}
	\end{eqnarray}
where $A$ and $\delta$ are integration constants. Here $A$ is related to the energy integral by $H=E=\frac{\frac{1}{2}w A^2}{1+\lambda A^2}$.

In Figure 3 we demonstrate both the harmonic oscillator Eq. (\ref{modeq1}) and the nonlinear oscillator Eq. (\ref{mod94}) exhibiting periodic oscillations with the same amplitude for identical initial conditions. However, the nonlinear oscillator Eq. (\ref{mod94}) exhibits amplitude dependent frequency of oscillations.  By changing the initial conditions, the amplitude of the oscillations also gets changed. Depending on the value of the amplitude $A$, the frequency $\Omega$ of the nonlinear oscillator (\ref{mod94}) also changes. Further, this is also confirmed by the solution given in (\ref{ml_soln}), where the amplitude $A$ is related with the frequency $\Omega$ through the expression $\Omega^2=\frac{\omega^2}{1+\lambda A^2}$.
\begin{figure*}	
\centering
\resizebox{0.5\textwidth}{!}{
			\includegraphics{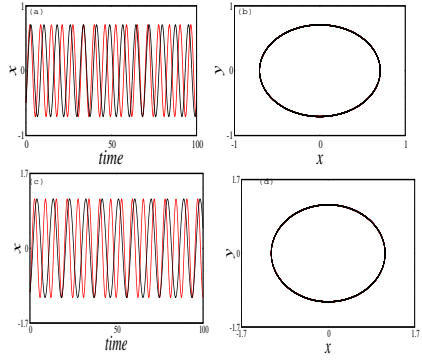}}
		\caption{Time series plot of equations (\ref{modeq1})  and (\ref{mod94}) (red and black) exhibiting periodic oscillations with same amplitude but with different frequencies for different initial conditions (a) $x(0)=u(0)=-0.5$ and $y(0)=v(0)=0.5$ and (c)  $x(0)=u(0)=-1$ and $y(0)=v(0)=0.5$. Corresponding phase space portraits are given in (b) and (d) for $\lambda= \omega=1.0$.}  
	\end{figure*}

 \subsubsection{Example: 2-coupled nonlinear Mathews-Lakshmanan (ML) oscillator equation}
Here, we consider a system of 2-coupled nonlinear ML oscillator equation whose equation of motion is given by
	\begin{eqnarray}
	\dot{x}_1=f w_1 y_1,~~
	\dot{x}_2=f w_2 y_2,~~
	\dot{y}_1=-f w_1 x_1,~~
	\dot{y}_2=-f w_2 x_2,	
	\label{eq:}
	\end{eqnarray}
where $f=[1+\lambda (x_1^2+x_2^2+y_1^2+y_2^2)]^{-\frac{1}{2}}$. Rewriting the above four first order ODEs into two coupled second order ODEs in the variables $x_1$ and $x_2$, we find
		\begin{eqnarray}
		(1+\lambda(x_1^2+x_2^2))\ddot{x}_1-\lambda x_1(\dot{x}_1^2+\frac{w_1^2}{w_2^2}\dot{x}_2^2)+w_1^2x_1=0,\nonumber\\
		(1+\lambda(x_1^2+x_2^2))\ddot{x}_2-\lambda x_2(\frac{w_2^2}{w_1^2}\dot{x}_1^2+\dot{x}_2^2)+w_2^2x_2=0,
		\label{eq:}
		\end{eqnarray}
	where $\lambda$, $w_1$ and $w_2$ are arbitrary parameters.  
	\par Following the procedure given in the previous section, we find the general solution in the form
	\begin{eqnarray}
	x_1=\sqrt{H_1}\sin((1+\lambda(H_1+H_2)^{(-1/2)})w_1  t+\delta_1),\nonumber\\
	y_1=\sqrt{H_1}\cos((1+\lambda(H_1+H_2)^{(-1/2)})w_1 t+\delta_1),\nonumber\\
	x_2=\sqrt{H_2}\sin((1+\lambda(H_1+H_2)^{(-1/2)})w_2 t+\delta_2),\nonumber\\
	y_2=\sqrt{H_2}\cos((1+\lambda(H_1+H_2)^{(-1/2)})w_2 t+\delta_2),
	\label{eq:}
	\end{eqnarray}
where $H_i$ and $\delta_{i},~i=1,2,3,4$, are integration constants. One important point we wish to note here is that the general solution for the coupled ML oscillators has been given for the case $w_1 \neq w_2$. In Ref. \cite{sir1}, the authors have generalized the one dimensional ML oscillator to two dimensions as well as $n$-dimensions by considering the parameter $w$ as same in all the dimensions.

 \subsubsection{Example: $2N$-coupled nonlinear ML oscillator equation}
The equation of motion for the $2N$-coupled nonlinear ML oscillator can be written as
\begin{eqnarray}
	\label{mod8dfs1}
	\dot{x}_{i}=f(\overline{x},\overline{y})\omega_{i}y_{i},~~
	\dot{y}_{i}=-f(\overline{x},\overline{y})\omega_{i}x_{i},\;\;\;\;i=1,2,\cdots,N,
	\end{eqnarray}
where $f(\overline{x},\overline{y})=f=[1+\lambda (x_1^2+x_2^2+\cdots+x_N^2+y_1^2+y_2^2+\cdots+y_N^2)]^{-\frac{1}{2}}$. Rewriting the above $2N$ first order ODEs into $N$-coupled second order ODEs in the variables $x_1,x_2,\cdots,x_N$, we find
\begin{eqnarray}
\label{cmod96}
(1+ \lambda \sum_{j=1}^{N}x_j^{2})\ddot{x}_i- \lambda x_i  \sum_{j=1}^{N}\frac{w_i^2}{w_j^2}\dot{x}_j^{2}+\omega_i^{2}x_i=0,
\end{eqnarray}
where $\lambda$ and $\omega_i^2$ are arbitrary constants. Choosing $f(\overline{x},\overline{y})=f(\overline{H})=(1+\lambda \sum_{i=1}^{N}H_{i})^{-1/2}$, we obtain the general solution of $(\ref{cmod96})$ in the form
	\begin{eqnarray}
	\label{mod98}
	x_{i}(t)&=&\sqrt{H_{i}}\sin\Bigg(\bigg(1+\lambda\sum_{j=1}^{N}H_{j}\bigg)^{-1/2}\omega_{i}t+\delta_{i}\Bigg),\nonumber \\
	y_{i}(t)&=&\sqrt{H_{i}}\cos\Bigg(\bigg(1+\lambda\sum_{j=1}^{N}H_{j}\bigg)^{-1/2}\omega_{i}t+\delta_{i}\Bigg),\hspace{0.1cm}i=1,2,...,N,
	\end{eqnarray}
where  $H_{i}$ and $\delta_i$ are the integration constants. 

\section{Conclusion}
    In this work, we have discussed a method of generating isochronous and non-isochronous nonlinear oscillators from the integrals of the simple harmonic oscillator equation. By considering the integrals to be in the same form for both the linear and nonlinear oscillators, we have identified the nonlinear oscillators that posses either isochronous and non-isochronous property (involving harmonic functions). We then considered two coupled first order ODEs and discussed the method of constructing the solution for both the cases. In each case we have also demonstrated the theory with examples. We then extended the theory to $2N$ coupled first order ODEs. We have also derived the general solution and presented the explicit forms of integrals for both the cases. We have also given examples for each of the cases. As far as the isochronous systems are concerned, the value of the exponent $q$ for the function $f$ in Sec. 3 decides the nature of the solution which will be either periodic or damped oscillations. The general solution and the integrals for the $2N$-coupled ML oscillator system which possesses the non-isochronous property are also presented for the first time in the literature.

\section{Acknowledgments}
RMS is funded by the Center for Nonlinear Systems, Chennai Institute of Technology, India, vide funding number CIT/CNS/2021/RP-015.
V.K.C. thanks DST, New Delhi for computational facilities under the DST-FIST programme (Grant No. SR/FST/PS- 1/2020/135) to the Department of Physics. V.K.C. is also supported by the SERB-DST-MATRICS Grant No. MTR/2018/000676. The work of M.S forms part of a research project sponsored by National Board for Higher Mathematics (NBHM), Government of India under Grant No. 02011/20/2018NBHM(R.P)/R\&D II/15064 and M.L. is supported by a DST-SERB National Science Chair.


\end{document}